\documentclass[preprint,12pt]{elsarticle}
\journal{Computer Physics Communications}

\usepackage[T1]{fontenc} 

\usepackage[utf8]{inputenc}
\usepackage{url}
\usepackage{graphicx}
\usepackage{axodraw2}

\newcommand{\be}{\begin{equation}}
\newcommand{\ee}{\end{equation}}
\newcommand{\bea}{\begin{eqnarray}}
\newcommand{\eea}{\end{eqnarray}}
\newcommand{\dd}{\mbox{d}}
\newcommand{\pa}{\partial}
\newcommand{\al}{\alpha}

\begin{document}

\begin{frontmatter}

\title{
\normalfont
\vskip-1cm{\baselineskip14pt
  \begin{flushleft}
      \normalsize SFB/CPP-14-60
  \end{flushleft}}
  \vskip1.5cm
  \boldmath 
FIRE5: a C++ implementation of Feynman Integral REduction}

\author[SRCC]{A.V.~Smirnov\corref{cor1}}
\ead{asmirnov80@gmail.com}


\cortext[cor1]{Corresponding author}

\address[SRCC]{Scientific Research Computing Center, Moscow State University, 119992\\ Moscow, Russia}

\begin{abstract}
In this paper the {\tt C++} version of {\tt FIRE} is presented --- a powerful program performing
Feynman integral reduction to master integrals. All previous versions used only
{\tt Wolfram Mathematica}, the current version mostly uses {\tt Wolfram Mathe\-ma\-ti\-ca} as a front-end.
However, the most complicated part, the reduction itself can now be done by {\tt C++}, which significantly improves the performance and allows one to reduce Feynman integrals
in previously impossible situations.
\end{abstract}
\end{frontmatter}
\newpage

{\bf PROGRAM SUMMARY}

\vspace{1cm}

\begin{small}
\noindent
{\em Manuscript Title:} FIRE5: a C++ implementation of Feynman Integral REduction\\
{\em Authors:} A.V. Smirnov\\
{\em Program title:} FIRE5\\
{\em Licensing provisions:} GPLv2\\
{\em Programming language:} {\tt Wolfram Mathematica} 6.0 or higher, {\tt C++}\\
{\em Computer(s) for which the program has been designed:} starting from a desktop PC\\
{\em Operating system(s) for which the program has been designed:} Linux 64bit\\
{\em RAM required to execute with typical data:} depends on the complexity of the problem  \\
{\em Has the code been vectorized or parallelized?:} yes\\
{\em Number of processors used: } the program has been tested on computers with up to 32 cores, but it is not a requirement \\
{\em Supplementary material:} The article, install instructions, http://science.sander.su
{\em CPC Library Classification:} 4.4 Feynman diagrams, 4.8 Linear Equations and Matrices, 5 Computer Algebra, 20 Programming and Publication Practice \\
{\em External routines/libraries used:} {\tt Wolfram Mathematica} [1], {\tt Snappy} [2],  {\tt KyotoCa\-binet} [3], {\tt Fermat} [4], {\tt LiteRed} [5], \\
{\em Nature of problem:}
Reducing Feynman integrals to master integrals can be treated as a task to solve
a huge system of sparse linear equations with polynomial coefficients.\\
{\em Solution method:}
Since the matrix of equations is very specific, none of standard methods of solving linear 
equations can be applied efficiently. The program approaches solving those equations 
with a special version of Gauss elimination.
The data preparation and result analysis is performed in {\tt Wolfram Mathematica} [1],
but the main reduction procedure is written in {\tt C++};
{\tt FIRE} compresses data with the use of the {\tt Snappy} [2] library,
stores it on disk with the use of the {\tt KyotoCabinet} [3] database engine,
and performs algebraic simplifications with the {\tt Fermat} [4] program.
The external package {\tt LiteRed} [5] can be used to produce additional
rules for reduction. \\
{\em Restrictions:} The complexity of the problem is mostly restricted
by CPU time required to perform the reduction of integrals and the available RAM.
The program has the following limits: maximal number of indices = 21, maximal number of positive indices = 15, 
maximal number of non-trivial sectors = $128\times256-3=32765$ (global symmetries decrease the number of sectors, indices that cannot be positive 
do not double the number of sectors). \\
{\em Running time:} depends on the complexity of the problem\\
{\em References:} 
{\\}
[1] \url{http://www.wolfram.com/mathematica/}, commercial algebraic software; 
{\\} [2] \url{http://code.google.com/p/snappy/}, open source;
{\\} [3] \url{http://fallabs.com/kyotocabinet/}, open source;
{\\} [4] \url{https://home.bway.net/lewis/}, free--ware with some restrictions for organizations;
{\\} [5] \url{http://www.inp.nsk.su/~lee/programs/LiteRed/}, open source.

\end{small}

\newpage

\section{Introduction}

In modern elementary particle physics one often needs to 
evaluate thousands and millions of Feynman integrals. 
An already classical approach is to apply the so-called
\textit{integration by parts (IBP) relations}~\cite{Chetyrkin:1981qh}  
(see Chapter~6 of \cite{Smirnov:2013ym} for a recent review)
and reduce all integrals to a smaller set,
the \textit{master integrals}\footnote{It has been shown in \cite{Smirnov:2010hn} that the number of master integrals is
always finite, so theoretically, this approach should be successful.}. 

There are multiple programs performing the task of Feynman integral reduction,
one of those presented by the author of this paper a few years ago.
The initial version of {\tt FIRE}~\cite{Smirnov:2008iw,Smirnov:2013dia} was written in 
{\tt Wolfram Mathematica}.
For other public products see
\cite{Anastasiou:2004vj,Studerus:2009ye,vonManteuffel:2012np,Lee:2008tj,Lee:2012cn,Lee:2013mka}

The goal of this paper is to present the {\tt C++} version of {\tt FIRE} --- a powerful program for Feynman integral reduction.
Recently {\tt FIRE} was able to perform a reduction with about 3 billion integrals involved.
It was successfully applied in \cite{Baikov:2009bg,Lee:2010cga,Lee:2010ik,Lee:2011jf,Lee:2011jt,Eden:2012fe,
Lee:2013sx,Drummond:2013nda,Grigo:2013rya,Henn:2013tua,Henn:2013woa,
Bern:2013uka,Henn:2013nsa,Henn:2014lfa,Caola:2014lpa} as well as in several pending projects.

\section{Basic definitions}

This section almost entirely copies the similar section from the previous paper on {\tt FIRE4}.
However, the following definitions should be given to make the paper
self-consistent.

Let us consider a family  of Feynman integrals as
functions of $n$ integer variables (indices),
\bea
  F(a_1,\ldots,a_n) &=&
  \int \cdots \int \frac{\dd^d k_1\ldots \dd^d k_h}
  {E_1^{a_1}\ldots E_n^{a_n}}\,,
  \label{eqbn-intr}
\eea
where the denominator factors $E_i$ are linear functions with respect to
scalar products of loop momenta $k_i$ and external momenta $p_i$, and dimensional regularization with
$d=4-2\epsilon$ is applied.

The integration by parts relations~\cite{Chetyrkin:1981qh} 
\bea \int\ldots\int \dd^d k_1 \dd^d k_2\ldots
\frac{\pa}{\pa k_i}\left( p_j \frac{1}{E_1^{a_1}\ldots E_n^{a_n}}
\right)   =0   \label{RR-intr}
\eea
can be rewritten in the following form:
\begin{equation}
\sum \al_i F(a_1+b_{i,1},\ldots,a_n+b_{i,n})
=0\,.
\label{IBP-intr}
\end{equation}
where $b_{i,j}\in \{-1, 0, 1\}$ and $\al_i$ are linear functions of $a_j$.

A classical approach is to separate all possible sets of indices into so-called sectors. Choosing a sector (one out of $2^n$) defines
for each index $a_i$ whether it is positive or non-positive. 
In fact, there are less than $2^n$ sectors --- indices corresponding to irreducible numerators are always non-positive.
A \textit{corner integral} in a sector is the one with indices equal to $0$ or $1$; each sector has a unique corner integral.

A sector is said to be \textit{lower} than another sector if all indices of the corner integral in the first one are smaller that corresponding
indices of the corner integral in the second one. Normally one tries to reduce Feynman integrals to those corresponding to lower sectors.
The reason for such a choice is that positive shifts always come 
with multiplication by the corresponding index, therefore relations written in sectors with negative values of indices do not 
depend on integrals with positive values of those indices. Moreover, integrals are simpler if more indices are non-positive.

The complexity of each integral corresponding to a given family (\ref{eqbn-intr}) is basically defined by 
the number of positive indices, and then the two non-negative numbers $N_+=\sum_{i\in \nu_{+}} (a_i-1)$ 
(the number of \textit{dots})
and
$N_-=-\sum_{i\in \nu_{-}} a_i$, where $\nu_+$ ($\nu_-$) are sets of positive (negative) indices.

A sector is called \textit{trivial} if all integrals corresponding to sets of indices in this sector are equal to zero. The sector 
with all non-positive indices is always trivial. The conditions determining whether a sector is trivial are called
\textit{boundary conditions}.

A \textit{Laporta algorithm}~\cite{Laporta:2001dd} in a given sector is solving IBPs with a Gauss elimination after choosing an ordering.
The ordering choice and all details of the algorithm can be modified by the algorithm implementer.

\section{Installation}

There are two ways to get {\tt FIRE5} --- one can either download a binary package from \url{http://git.sander.su/fire/downloads} or build it from sources.
The binary package is not guaranteed to work at a particular machine.
To build {\tt FIRE5} from sources one has to 

1) clone it with {\tt git} (it should be done in a folder that has no space symbols in its full path)

{\tt git clone https://bitbucket.org/feynmanIntegrals/fire.git}

2) install the libraries {\tt FIRE5} requires system-wide or build them from sources. 
{\tt FIRE} comes with the database engine {\tt kyotocabinet 1.2.76} and compression library {\tt snappy 1.1.1}.

{\tt make dep}

3) build the {\tt FIRE} binaries

{\tt make}

Do not forget to change the current directory to {\tt fire/FIRE5} before giving those commands.
To build {\tt FIRE} and the depending libraries faster, add {\tt -j n} to the {\tt make} commands
as usual, where {\tt n} is the number of kernels on the computer in use.

To test whether the binaries were compiled properly run

{\tt make test}

If one is compiling {\tt FIRE} at a cluster and is going to use only the {\tt C++ FIRE}, it can be compiled it with

{\tt make nomath}

To get the latest development version of {\tt FIRE}, one should switch to the {\tt dev} branch and compile {\tt FIRE} again:

{\tt
git checkout --track -b dev origin/dev
}

{\tt
make
}

{\tt FIRE5} comes with the external program {\tt fermat} performing polynomial algebra. The only copyright statement of {\tt fermat} on the web-page says 
``Fermat is free--ware. Those who have grant money are requested to contribute \$60 per installed machine.'' 
Those who are not sure how this applies to them should not use {\tt FIRE} before consulting with the {\tt fermat} author (\url{https://home.bway.net/lewis/}).
The version shipped with {\tt FIRE} is the 64-bit version 5.16 for {\tt Linux}. 

As for other operating systems, {\tt FIRE} works at {\tt Linux 32bit}, but one should replace the {\tt fermat} binary.
It should also be possible to build {\tt FIRE} at {\tt MAC OS X}, but there might be some problems during the installation.
If one is interested in building it at {\tt MAC OS X} please contact the author of the paper, and perhaps we will work out 
the best way to do it, so that these instructions are included in the next release.
And I do not think it is possible to make the {\tt C++ FIRE} work under {\tt Windows}.

\section{Usage of {\tt FIRE}}

All the syntax of {\tt FIRE5} is made backward-compatible with the syntax of the previous versions. This might look old-fashioned and the syntax is sometimes weird, however there are too many people who have been using {\tt FIRE} for quite some time.

\subsection{Preparing a start file}

The first thing one needs to do to work with {\tt FIRE} is to prepare a start file containing the information on a given diagram. To do this, one needs to load {\tt FIRE}, define the internal (loop) and external momenta, the propagators, the replacements (values of kinematic invariants). If one is going to use {\tt C++}, 
it is important to have all variables corresponding to kinematic invariants to have no capital letters in their names (due to the restrictions of the {\tt fermat} program).

If one is loading {\tt FIRE} from {\tt Ma\-the\-ma\-tica}, it should either be loaded with

{\tt SetDirectory[<path to FIRE>]; Get["FIRE5.m"];} 

or 

{\tt FIREPath=<path to FIRE>; Get[FIREPath<>"FIRE5.m"];}.

Do not load {\tt FIRE} by simply specifying a full path to it, might fail to work properly.

The basic syntax to create a start file is the following: one has to give proper values to the following variables:

\begin{itemize}
 \item {\tt Internal} --- the list of internal momenta, for example, \{k\};
 \item {\tt External} --- the list of external momenta, for example, \{p1, p2, p4\};
 \item {\tt Propagators} --- the list of propagators, for example, \{-k$^2$, -(k + p1)$^2$, -(k + p1 + p2)$^2$, -(k + p1 + p2 + 
       p4)$^2$\};
 \item {\tt Replacements} (optional) --- the list of replacement rules for kinematic invariants, for example, \{p1$^2$ -> 0, p2$^2$ -> 0, p4$^2$ -> 0, p1 p2 -> s/2, 
   p2 p4 -> t/2\};
 \item {\tt RESTRICTIONS} (optional) --- list of boundary conditions. For example if this list has an element {\tt \{-1, -1, -1, 0\}}, this means that the integrals are equal to zero if the first three indices are non-positive;
 \item {\tt SYMMETRIES} (optional) --- list of symmetries (permutations of indices not changing the integrals); in older versions one had to provide the whole symmetry group, but currently it is enough to provide the generators; for example,  if this list has an element {\tt \{3, 2, 1, 4\}}, this means that {\tt F[a,b,c,d]=F[c,b,a,d]}.
\end{itemize}

After executing these commands one should run {\tt PrepareIBP[]} and then {\tt Prepare{}} or, preferably, {\tt Prepare[AutoDetectRestrictions -> True]} 
to automatically detect boundary conditions (in both cases the conditions provided by the {\tt RESTRICTIONS} setting are taken into account).
However if one is planning to use {\tt LiteRed} \cite{Lee:2008tj,Lee:2012cn,Lee:2013mka} (see section \ref{LiteRed}), then {\tt FIRE} will read boundary conditions from {\tt LiteRed} output later,
so it is not obligatory to provide them at the current stage. 

In older versions one also had to give a value to the {\tt startinglist} variable to provide the list of IPBs.
It is still possible, however is not obligatory. If the {\tt startinglist} is skipped, the code
uses all IBP relations, where
one multiplies by an internal or external momenta and differentiates in an internal momenta, 
and then replacements are applied. Internally it uses the following construction:

{\tt
startinglist = 
  Flatten[Outer[(IBP[\#1, \#2] //.Replacements) \&, \\Internal, 
    Join[Internal, External]], 1];
}

Now a start file can be saved with {\tt SaveStart["filename"]}. It is recommended to quit the kernel afterwards.

For example, for a massless double box diagram one has:

\tt
FIREPath = <path to the folder with FIRE>;

Get[FIREPath<>"FIRE5.m"];
     
Internal = \{k1, k2\};

External = \{p1, p2, p3\};

Propagators = \{-k1$^2$, -(k1 + p1 + p2)$^2$, -k2$^2$, -(k2 + p1 + 
       p2)$^2$, -(k1 + p1)$^2$, -(k1 - k2)$^2$, -(k2 - p3)$^2$, -(k2 + 
       p1)$^2$, -(k1 - p3)$^2$\};
       
Replacements = \{p1$^2$ -> 0, p2$^2$ -> 0, p3$^2$ -> 0, p1 p2 -> s/2, \\
   p1 p3 -> t/2, p2 p3 -> -1/2 (s + t)\};       
       
PrepareIBP[];

Prepare[AutoDetectRestrictions -> True];

SaveStart["doublebox"];

Quit[];

\normalfont

As a result one has a file {\tt "box.start"} containing all the required information to proceed.

\subsection{Reduction in {\tt Wolfram Mathematica}}

To perform the reduction by {\tt Mathematica} (or to load tables at a later stage) one has to load
a start file. This is done by the {\tt LoadStart} command. For example, 

\tt
FIREPath = <path to the folder with FIRE>;

Get[FIREPath<>"FIRE5.m"];

LoadStart["doublebox", 1];

Burn[]
\normalfont

The second argument in {\tt LoadStart} is some positive integer number less than $1000$ that is assigned to the current family of Feynman integrals.
If the problem has multiple families, it is recommended to enumerate them and use different numbers for different families.

Now one can perform the reduction with

\tt

F[1, \{1, 1, 1, 1, 1, 1, 1, -1, -1\}]

\normalfont

Here {\tt F} is a call to perform the reduction, {\tt 1} is the same number as in {\tt LoadStart} and {\tt \{1, 1, 1, 1, 1, 1, 1, -1, -1\}} is the set of indices.
As a result one has something like

\begin{eqnarray}
\frac{3}{2} s G(1,\{1,1,1,1,1,1,1,-1,0\}) +\frac{1}{2} s t G(1,\{1,1,1,1,1,1,1,0,0\}) +\ldots \nonumber
\end{eqnarray}

In the resulting expression {\tt G} follows the same notation as {\tt F} but is treated as an irreducible (master) integral and does not lead to new reduction.

Alternatively one can use the {\tt EvaluateAndSave} command, which has two arguments --- the list of integrals that has
to be reduced and the file where to save tables. For example, to reduce two integrals by this command one should run

{\tt EvaluateAndSave[\{\{1, \{1, 1, 1, 1, 1, 1, 1, -1, -1\}\},\\ \{1, \{1, 1, 1, 1, 1, 1, 1, 0, -2\}\}\},"doublebox.tables"]}

This is the recommended way to run {\tt Mathematica FIRE} if
the reduction takes long enough, however it is even better to use the {\tt C++ FIRE} which is a direct 
replacement to the {\tt EvaluateAndSave} command.

In complicated situation {\tt Burn[]} might work slowly. Hence one can run
{\tt SaveData["filename"} after {\tt Burn[]}, then quit the kernel and later
load everything with {\tt LoadData["filename"]} (without {\tt LoadStart} and {\tt Burn}).

\subsection{Finding equivalents between master integrals}

As one can clearly see from the last example, the integrals that {\tt FIRE} claims to be irreducible can be symmetric and thus equivalent.
Normally this happens with integrals from different sectors of same level --- {\tt FIRE} cannot find equivalents between them during the reduction
run. However, there is an alternative method that can find the missing relations. 

If in the last example one gives the {\tt MasterIntegrals[]} command, the result is:

{\tt
\noindent
\{\{1, \{0, 0, 0, 0, 1, 1, 1, 0, 0\}\}, \{1, \{0, 0, 1, 1, 1, 1, 0, 0, \\
   0\}\}, \{1, \{0, 0, 1, 1, 1, 1, 1, 0, 0\}\}, \{1, \{0, 1, 1, 0, 0, 1, 0, 0,
    0\}\}, \{1, \{0, 1, 1, 0, 1, 1, 1, 0, 0\}\}, \{1, \{1, 0, 0, 1, 0, 1, 0, 
   0, 0\}\}, \{1, \{1, 0, 0, 1, 1, 1, 1, 0, 0\}\}, \{1, \{1, 1, 0, 0, 0, 1, 1,
    0, 0\}\}, \{1, \{1, 1, 0, 0, 1, 1, 1, 0, 0\}\}, \{1, \{1, 1, 1, 1, 0, 0, 
   0, 0, 0\}\}, \{1, \{1, 1, 1, 1, 1, 1, 1, 0, 0\}\}, \{1, \{1, 1, 1, 1, 1, 1,
    1, -1, 0\}\}\}
}

It is easy to see that some of the integrals in lower sectors are equivalent.
This can be found with 

\tt

Internal = \{k1, k2\};

External = \{p1, p2, p3\};

Propagators = \{-k1$^2$, -(k1 + p1 + p2)$^2$, -k2$^2$, -(k2 + p1 + 
       p2)$^2$, -(k1 + p1)$^2$, -(k1 - k2)$^2$, -(k2 - p3)$^2$, -(k2 + 
       p1)$^2$, -(k1 - p3)$^2$\};
       
Replacements = \{p1$^2$ -> 0, p2$^2$ -> 0, p3$^2$ -> 0, p1 p2 -> s/2, \\
   p1 p3 -> t/2, p2 p3 -> -1/2 (s + t)\};       

FindRules[MasterIntegrals[]]   
   
\normalfont

or saved to a file with

{\tt
WriteRules[MasterIntegrals[], \\ FIREPath <> "examples/doublebox"];
}

The command {\tt MasterIntegrals} by itself is a replacement for the old {\tt GetII /\& IrreducibleIntegrals[]}
and produces the list of integrals that {\tt FIRE} found to be irreducible.

Then if one loads the rules after {\tt Burn[]} with

{\tt
LoadRules[FIREPath <> "examples/doublebox", 1];
}

\noindent and performs the reduction again, there will be only $8$ master integrals instead of $12$.

The file with rules consists of {\tt Mathematica} rules, however the right-hand side should be not a linear combination,
but a list of pairs --- coefficients and integrals (in order to simplify parsing in {\tt C++}). In our examples the file has $4$ lines, and they look like

{\tt G[1, \{0, 0, 1, 1, 1, 1, 1, 0, 0\}] -> \\\{\{1, G[1, \{1, 1, 0, 0, 1, 1, 1, 0, 0\}]\}\};}

{\tt G[1, \{1, 0, 0, 1, 1, 1, 1, 0, 0\}] -> \\\{\{1, G[1, \{0, 1, 1, 0, 1, 1, 1, 0, 0\}]\}\};}

{\tt G[1, \{1, 1, 0, 0, 0, 1, 1, 0, 0\}] -> \\\{\{1, G[1, \{0, 0, 1, 1, 1, 1, 0, 0, 0\}]\}\};}

{\tt G[1, \{1, 0, 0, 1, 0, 1, 0, 0, 0\}] -> \\\{\{1, G[1, \{0, 1, 1, 0, 0, 1, 0, 0, 0\}]\}\};}

If one knows some relations between master integrals, they can also be added manually to the file with rules \cite{Smirnov:2013dia}.

\subsection{{\tt C++} reduction}

The main power of {\tt FIRE} comes with the {\tt C++} part of the program.
It uses the same start and rules files as input and saves table files as a result 
so that they can be loaded into {\tt Mathematica} afterwards. 

To run the {\tt C++} version, one has to create a configuration file (with the {\tt config} extension).
Let us illustrate the syntax with this example:

\tt
$\#$threads           4

$\#$variables         d, s, t

$\#$start

$\#$folder            examples/

$\#$problem           1 doublebox.start

$\#$integrals         doublebox.m

$\#$output            doublebox.tables
\normalfont

The number of space symbols in this file in unimportant. Each line should start from $\#$.
The $\#${\tt threads} command specifies the number of threads that can work in parallel, 
$\#${\tt variables} is important and should list all variables that can appear, if variables
are set incorrectly, the reduction will freeze. 
These (and some other optional directives listed in section~\ref{config}) should be followed by the $\#${\tt start} command.

The next part of the configuration file is related to a specific diagram.
$\#${\tt folder} is optional and provides a path to a folder where other files can reside. 
If the folder instruction is missing, the paths are considered absolute or from the ``current'' directory.
$\#${\tt problem} followed by the number corresponding to the current family of Feynman integrals and the path to a start file. 
$\#${\tt integrals} points to a file 
containing a list of integrals that have to be reduced and $\#${\tt output} points to a file where
the resulting tables are to be saved. The input is a {\tt Mathematica} list of pairs, for example,
{\tt \{\{1, \{1, 1, 1, 1, 1, 1, 1, -1, -1\}\}, \{1, \{1, 1, 1, 1, 1, 1, 1, 0, -2\}\}\}}.

Now one can launch the reduction with 

{\tt bin/FIRE5 -c examples/doublebox}

It should be stated once more that this reduction does not need {\tt Ma\-the\-ma\-ti\-ca}. As a result one gets a file with tables.
While the {\tt Mathematica} reduction took $1453$ seconds on my laptop ($1386$ seconds with rules), the $\tt C++$ reduction took only $92$ seconds.
And for diagrams having more loops the difference can be much more significant.

This reduction can also make use of the rules previously saved in a file. To do that, add the following line to the configuration file:

{\tt $\#$rules         doublebox.rules}

This example also shows that the choice of master integrals might be different in different versions of {\tt FIRE}.
The {\tt C++ FIRE} tends to prefer integrals without negative indices, because it is impossible to find equivalents 
in case of irreducible numerators. However in this particular example the choice
{\tt \{\{1, \{1, 1, 1, 1, 1, 1, 2, 0, 0\}\}\}} instead of {\tt \{\{1, \{1, 1, 1, 1, 1, 1, 1, -1, 0\}\}\}} makes the answer
more complicated. One can make {\tt FIRE} do another choice by means of the preferred list. To do that, one should add

{\tt $\#$preferred         doublebox.preferred}

to the configuration file, where this file contains a {\tt Mathematica} list of preferred master integrals.
In this example this file contains one element,  {\tt \{\{1, \{1, 1, 1, 1, 1, 1, 1, -1, 0\}\}\}}.
Now if one runs 

{\tt bin/FIRE5 -c examples/doubleboxrp}

the job will take only $62$ seconds.

\subsection{Loading the tables}

The tables obtained with the {\tt C++} reduction (or with the {\tt EvaluateAndSave} command in {\tt Mathematica}) can be loaded into {\tt Mathematica}.

\tt
FIREPath = <path to the folder with FIRE>;

Get[FIREPath<>"FIRE5.m"];

LoadStart["doublebox", 1];

Burn[];

LoadTables["doublebox.tables"];
\normalfont

Now a call to {\tt F} will not need a new reduction, but take the result from tables.
However, sometimes this can be slow because {\tt FIRE} tries to factorize coefficients in order to present a 
nice-looking result. This behavior can be switched off by {\tt FactorCoefficients = False}. 

\subsection{Using {\tt LiteRed} rules in {\tt Mathematica}}

\label{LiteRed}

{\tt LiteRed} is a program by Lee \cite{Lee:2008tj,Lee:2012cn,Lee:2013mka} that aimes on solving the IBP relations
before the substitution of indices. If the program succeeds, the reduction can be significantly more effective and fast.
But even if reduction rules cannot be constructed everywhere, it makes sense to use a part of them in order to 
speed up the reduction, or at least the information on boundary conditions and symmetries.

Currently {\tt FIRE} comes shipped with the version {\tt 1.6} of {\tt LiteRed}. If one wishes to upgrade to the most recent
version of {\tt LiteRed} without waiting for an update of {\tt FIRE}, 
it can be downloaded from \url{http://www.inp.nsk.su/~lee/programs/LiteRed/}.

{\tt LiteRed} starts with finding symmetric sectors. Two sectors are considered symmetric, 
if integrals from one of the sectors can be represented as linear combinations of integrals in the other. {\tt LiteRed} is quite effective
in finding symmetries between sectors and this procedure does not take much time. So even if one is working with 
complicated diagrams where the solution of IBPs cannot be performed, the symmetries should be taken into account.

{\tt FIRE} is capable of loading {\tt LiteRed} rules. To do this, one has to construct them first using {\tt LiteRed}. 
There are multiple examples shipped with the {\tt LiteRed} package, but in the contextity of {\tt FIRE} it is convenient to do with

\tt

FIREPath = <path to the folder with FIRE>;

SetDirectory[FIREPath <> "extra/LiteRed/Setup/"];

Get["LiteRed.m"];

SetDim[d];

Declare[\{l, r, p, q\}, Vector];

p$\cdot$p = 0; q$\cdot$q = 0; p$\cdot$q = -1/2;

NewBasis[v2, \{-sp[l - r], -sp[l], -sp[r], -sp[p - l], -sp[q - r], -sp[p - l + r],  -sp[q - r + l]\}, \{l, 
  r\}, \\ Directory -> FIREPath <> "temp/v2.dir"];
  
GenerateIBP[v2];

AnalyzeSectors[v2, \{0, \_\_\}];

FindSymmetries[v2,EMs->True];

SolvejSector /@ UniqueSectors[v2];

DiskSave[v2];

Quit[];

\normalfont

Here a two-loop non-planar massless vertex figure is considered.
There is a different syntax in {\tt LiteRed} when compared with {\tt FIRE}.
{\tt SetDim[d]} chooses {\tt d} as dimension of space-time;
{\tt Declare[\{l, r, p, q\}, Vector]} defines the listed variables as momenta;
the centered dot stands for the scalar product and is equivalent to {\tt sp} 
(in case of one argument {\tt sp} is treated as the scalar square).
{\tt NewBasis} creates a new {\tt LiteRed} bases has the following arguments:
a label for the current family of Feynman integrals, a list of propagators,
a list of loop momenta and some options such as directory choice.

However, it is possible to translate one input into another. The code above is equivalent 
(both can be used but the second variant allows one to provide input only in the style of {\tt FIRE}) to

\tt

FIREPath = <path to the folder with FIRE>;

SetDirectory[FIREPath <> "extra/LiteRed/Setup/"];

Get["LiteRed.m"];

Get[FIREPath <> "FIRE5.m"];

Internal = \{l, r\};

External = \{p, q\};

Propagators = 
  (-Power[$\#\#$, 2]) \& /@ \{l - r, l, r, p - l, \\ q - r, p - l + r, q - r + l\};

CreateNewBasis[v2, Directory -> FIREPath <> "temp/v2.dir"];
  
GenerateIBP[v2];

AnalyzeSectors[v2, \{0, \_\_\}];

FindSymmetries[v2,EMs->True];

SolvejSector /@ UniqueSectors[v2];

DiskSave[v2];

Quit[];

\normalfont

Note: {\tt CreateNewBasis} is a not a command of {\tt LiteRed}, it is a command in {\tt FIRE} that translates {\tt FIRE} input into {\tt LiteRed} input.

The {\tt AnalyzeSectors} command finds trivial sectors, the {\tt FindSymmetries} finds symmetric sectors,
{\tt SolvejSector} builds reduction rules in sectors. 
{\tt SolvejSector} might take too long or even fail to complete in complex situations,
but even the usage of {\tt AnalyzeSectors} and {\tt FindSymmetries} can speed up the reduction
significantly.

As a result one gets a folder with many files containing information on the diagram.
Now one is ready to load those rules into the {\tt Mathematica FIRE}. Suppose a start file already exists
(this example among others can be found in the examples folder). Then one can do the following:

\tt

FIREPath = <path to the folder with FIRE>;

Get[FIREPath <> "FIRE5.m"];

LoadStart[FIREPath <> "examples/v2", 2];

LoadLRules[FIREPath <> "temp/v2.dir", 2];

Burn[];

\normalfont

The {\tt LoadLRules} command reads the directory with {\tt LiteRed} rules and loads 
everything it can use out of there. Then the reduction can be performes as before: {\tt F[2, \{-1, 1, 1, 1, 1, 1, 1\}]}.

\subsection{Using {\tt LiteRed} rules in {\tt C++}}

As mentioned before, the {\tt C++} reduction is much faster than the {\tt Mathema\-tica} one.
However the files created by {\tt LiteRed} are in such a format, that they cannot be used directly
in the {\tt C++} version and hence a convertion is necessary:

\tt

FIREPath = <path to the folder with FIRE>;

Get[FIREPath <> "FIRE5.m"];

LoadStart[FIREPath <> "examples/v2"]; 

TransformRules[FIREPath <> "temp/v2.dir", FIREPath <> \\ "examples/v2.lbases", 2];

SaveSBases[FIREPath <> "examples/v2"];

\normalfont

The start file is loaded without a diagram number here.
As a result of the {\tt TransformRules} the files in the {\tt "temp/v2.dir"} folder 
are transformed and saved in {\tt "v2.lbases"} file.
The number {\tt 2} corresponds to the current family of Feynman integrals, it is stored inside the lbases file and should
be also used when the file is being loaded later. The {\tt SaveSBases} command
saves information similar to the start file in {\tt "v2.sbases"}.
It is important to use the sbases file afterwards instead of the start file since
it stores some orderings that were used by {\tt LiteRed} and so should be used by {\tt FIRE} as well.

Afterwards a configuration file for the {\tt C++} reduction might look like:

\tt
$\#$threads           4

$\#$variables         d

$\#$start

$\#$folder            examples/

$\#$problem           2 v2.sbases

$\#$lbases            v2.lbases

$\#$integrals         v2.m

$\#$output            v2.tables
\normalfont

In this case the reduction works much faster than the one without {\tt LiteRed} rules.

\subsection{{\tt C++ FIRE} and databases}

To run an efficient reduction it is important to understand how the {\tt C++ FIRE} works with databases. 
It uses the {\tt kyotocabinet} database engine
to store them either in RAM or on disk. 
There is a database for each sector (storing relations) and one more ``common'' database 
(enumerating integrals).

If working in the ``RAM mode'', {\tt FIRE} keeps the common database and a number of sector databases (maximally equal to the number of threads) in RAM.
When the forward reduction in a sector is over, the database snapshot is saved to disk. It is be loaded back to RAM later at the substitutions stage.

If working in the ``disk mode'', all the databases are files on the disk. The disk access is much slower, however
if one is using a local disk, then normally the operating system efficiently caches those files with the use of RAM. 
When {\tt FIRE} finishes reduction in a sector, it explicitly instructs the operating system to stop caching that particular file,
hence the cache is spent only for databases that are currently needed.

It is hard to say which option is preferable, this depends on the configuration of the current computer in use.

There is also one more option that should be explained --- the size of the bucket array. The bucket array is some sort 
of index used by database engines. The bigger this index is, the more efficient the database is, however
the more RAM is needed to store the bucket array. The {\tt bucket} option is treated exponentially --- 
for example, the default value $20$ corresponds to about one million of entries in the bucket array.

For the hash database stored on disk (the sector databases are of this type) it is recommended that the number of entries in the database
should not be greater than $4$ times the number of entries in the bucket array. The tree database (the ``common database'' is of this type) 
and the RAM databases have lighter restrictions. 

The footprint of each record is $10$ bytes. For example, if one is using the ``disk mode'' and a bucket value of $27$ and is running $8$ threads at the same time,
then one needs about $2.5$GB for the bucket array.

\section{Additional options}

\subsection{Options in Mathematica}

{\tt FIRE} has some options that can be used in {\tt Mathematica}.

\begin{itemize}
 \item {\tt DatabaseUsage} --- a number from $0$ to $4$, by default equal to $0$. If it is different from $0$,
 {\tt FIRE} stores some information in a database. Hence in order for it to work, one needs working binaries.
 By default it uses the {\tt QLinkPath = bin/KLink} binary for data access and stores data in the {\tt temp} folder (all paths are 
 considered relative to the value of {\tt FIREPath}. If these files are on a network drive, 
 it might make sense to provide another {\tt DataPath};
 
 \item {\tt MemoryLimit} (in megabytes) --- an alternative to {\tt DatabaseUsage} that makes {\tt FIRE} increase the
 database usage (even if the initial value is {\tt 0}) upon reaching the memory limit;
 
 \item {\tt UsingFermat} --- a boolean option, {\tt False} by default. If set to {\tt True}, makes {\tt FIRE} use the external 
 {\tt fermat} program for algebraic simplifications. Since {\tt FIRE} comes now shipped with {\tt fermat}, there is probably 
 no need to edit the {\tt FermatPath = extra/ferl64/fer64} and the {\tt FLinkPath = bin/KLink} for the intermediate binary.
 
 \item {\tt FactorCoefficients} --- a boolean option, {\tt True} by default. If set to {\tt False}, {\tt FIRE} won't apply
 {\tt Factor} to coefficients when producing the final output;
 
\end{itemize}

\subsection{Config files for C++}

Let us explain in details the syntax of the configuration files (note: the order of some of the commands is important,
so to be safe it is recommended to keep the order as listed in this section).
\label{config}

\begin{itemize}

 \item {\tt $\#$fermat} (optional) --- the path to the {\tt fermat} binary. By default the binary shipped with {\tt FIRE} is used, but one might wish to change it, for example if {\tt Mac OS X} is used;
 \item {\tt $\#$threads} --- the number of threads launched for parallel reducing in sectors of same level; 
 \item {\tt $\#$fthread} (optional) --- the number of {\tt fermat} processes launched. By default it is equal to the number of threads, but it might make sense to increase it;
 \item {\tt $\#$variables} --- comma-separated list of variables used during the reduction;
 \item {\tt $\#$database} (optional) --- path to the place where {\tt FIRE} will store the data; by default it points to the {\tt temp} directory relative to the current folder, but this can be changed; it is especially important to be sure that this setting does not point to a network drive in case you do not use the {\tt memory} setting;
 \item {\tt $\#$bucket} (optional) --- an integer number equal to $20$ by default and related to the database engine; small values will make {\tt FIRE} auto-increase the bucket during the reduction, and this can slow things down, large values can make {\tt FIRE} use too much RAM; for complicated tasks consider a bucket value equal to $27$--$30$;
 \item {\tt $\#$memory} (optional) --- if this line exists, {\tt FIRE} will use ``disk mode'' instead of ``RAM mode'' --- it will store active databases in RAM; this setting makes {\tt FIRE} use more RAM, but it becomes less vulnerably to freezing because of network drives;
 \item {\tt $\#$start} --- just a command following the previous lines;
 \item {\tt $\#$folder} (optional) --- if this path is given, all following paths will be considered relative to this folder unless they are absolute paths (starting with {\tt /});
 \item {\tt $\#$problem} --- the instruction to load a start or sbases file; the syntax is {\tt $\#$problem pn path} or {\tt $\#$problem pn |maxpos|path} or {\tt $\#$problem pn |minpos,maxpos|path}; {\tt pn} here is the diagram number; if {\tt maxpos} is provided, then indices bigger than {\tt maxpos} cannot be positive; if {\tt minpos} is provided, then indices smaller than {\tt minpos} cannot be positive;
 \item {\tt $\#$rules} (optional) --- a command to load a file with rules for some integrals;
 \item {\tt $\#$lbases} (optional) --- a command to load with {\tt LiteRed} rules obtained by {\tt TransformRules};
 \item {\tt $\#$output} or {\tt $\#$masters} --- the path where {\tt FIRE} will store the resulting tables; if one chooses the {\tt $\#$masters} syntax, then {\tt FIRE} only aims at finding master integrals, and this can be much faster than the whole reduction; this might be needed to find master integrals, then one can use {\tt WriteRules} to find equivalents between them, so that afterwards the full reduction can be run with the use of those rules;
 \item {\tt $\#$preferred} (optional) --- this file can list integrals that will be preferred as master-integrals; this might be needed if you do not like the automatic choice; however, keep in mind that it is just a hint for {\tt FIRE};
 \item {\tt $\#$integrals} --- the file with integrals to be reduced.

\end{itemize}

The package contains a number of examples, that are described in the {\tt README} file.

\section{Optimization hints}

Here are some hints on how to optimize {\tt FIRE} performance:

\begin{itemize}
 \item Reduction should be performed by the {\tt C++} version of {\tt FIRE}, it is significantly faster; even a sample integral in the doubebox example in this paper 
 gave us a speedup by a factor of $20$, and the more complicated the reduction is, the more significant this speedup is; 
 \item Pay attention to the correct input. One should either list boundary conditions manually, or rely on {\tt AutoDetectRestrictions->True}, or use {\tt LiteRed} (at least the {\tt AnalyzeSectors} part); missing boundary conditions might reduce speed a lot;
 \item It is recommended to use {\tt LiteRed}, at least to reveal the symmetries with the {\tt FindSymmetries} command, however the {\tt SolvejSector} can decrease performance sometimes;
 \item Even after that one can have redundant master integrals; if the substitution stage runs too long, it makes sense to run the partial reduction only to reveal master integrals, find equivalents between them with the {\tt WriteRules} command and use them for the final reduction; some extra rules can be inserted manually;
 \item {\tt FIRE} normally prefers master integrals with denominators, but the command {\tt WriteRules} can find equivalents only for integrals without denominators, so to have them as masters (in cases when there are many masters in a single sector), one can use the {\tt preferred} option;
 \item Use parallelization via the {\tt threads} and {\tt fthreads} settings. Raise {\tt threads} as high as the number of kernels on your machine, but if that results in high RAM usage, {\tt threads} should be set to a smaller value but {\tt fthreads} can be left higher;
 \item It is highly recommended to have {\tt db} point at a directory located at a fast local hard disk or even an SSD (and for complicated cases one might need a TB of space or more), that is cached properly by means of the operating system; If it is not the case, the only option is to use the {\tt memory} option, however this results in greater RAM usage;
 \item Have enough RAM; swapping will most probably make reduction fail, but even if the RAM usage is far from the limit in the ``disk mode'', it is normally used for the cache (automatically, by means of the operating system), so a good amount of free RAM speeds the process a lot (without the {\tt memory} setting);
 \item Try to play with the {\tt bucket} setting --- a value from $27$ to $30$ is advised for complicated settings; a higher value speeds up the databases but gets the problem require more RAM (just for the ``bucket array'');
 \item If the {\tt Mathematica FIRE} takes too long to produce results loaded from tables, set {\tt FactorCoefficients=False}.
\end{itemize}

There are some more suggestions on what can be done if something is working to slow:

\begin{itemize}
 \item If {\tt Prepare} is working slow, then probably too many {\tt RESTRICTIONS} have been set for individual sectors. Try to use boundary conditions for sets of sectors or rely on {\tt AutoDetectRestrictions->True} or {\tt LiteRed} in detecting them; also the {\tt Parallel->True} option speeds up this procedure (use a number instead of {\tt True} to specify the number of cores);
 \item {\tt F} or {tt EvaluateAndSave} is working slow, then one should move to the {\tt C++} reduction;
 \item {\tt Burn[]} is working slow, then one can run it once and save results with {\tt SaveData[filename]}, then it can be loaded with {\tt LoadData[filename]} afterwards;
 \item The substitution stage of the {\tt C++} reduction (sectors counted upwards) is working too slow. It probably means that one has too many equivalent master integrals; equivalents should be revealed by {\tt LiteRed} or the {\tt FindRules} command after the run to detect irreducible integrals;
 \item The first part of the {\tt C++} reduction works too slowly; this may be due to different reasons but the first things to check are whether the database is located on a local disk (if not, the {\tt memory} option is obligatory), if one has enough RAM (and preferably enough RAM for caching the local databases), if one used a high enough number of threads; having located equivalent master integrals also helps at this stage; missing information on boundary conditions or symmetries slows things down; if the processor load is low, it mostly indicates disk or RAM problems.
\end{itemize}

\section{Conclusion}

The {\tt C++} version of {\tt FIRE} was presented--- a powerful program for Feynman integral reduction.
This version is backward-compatible with all the previous {\tt Mathematica} versions,
moreover the tasks are prepared in {\tt Mathematica} and the results can be read into 
{\tt Mathematica} as well. However the reduction itself can be now run in {\tt C++} with
a tremendous speedup when being compared with previous versions.

\section*{Acknowledgements}

This work was supported by DFG through SFB/TR 9.
I would like to thank R.~Lee, P.~Marquard, A.~Pak and M.~Tentyukov for useful advices and discussions
and V.~Smirnov and M.~Steinhauser for ongoing support as well as the careful reading of the draft of this paper.
 
\bibliographystyle{model1-num-names}
\bibliography{FIRE-new,asmirnov}
\end{document}